\documentclass[twocolumn]{aastex63}
\usepackage{newtxtext,newtxmath} 
\usepackage{graphicx}	
\usepackage{amsmath}	
\usepackage{amssymb}	
\usepackage{mathtools,float,hyperref}
\usepackage{color,ulem,epstopdf,placeins}
\usepackage[percent]{overpic}

\definecolor{darkgreen}{rgb}{0.2, 0.7, .2}

\newcommand{\z}{\color{orange}}
\newcommand{\x}{\sout}
\newcommand{\w}{\color{blue}}
\newcommand{\y}{\color{red}}

\accepted{\today}

\shortauthors{Kutra, Wu \& Qian}

\begin{document}

\title{Super-Earths and sub-Neptunes are Insensitive to Stellar Metallicity}

\correspondingauthor{Taylor Kutra}
\email{kutra@astro.utoronto.ca}

\author[0000-0002-7219-0064]{Taylor Kutra}
\affiliation{David A. Dunlap Department of Astronomy \& Astrophysics \\ 
University of Toronto}

\author[0000-0003-0511-0893]{Yanqin Wu}
\affiliation{David A. Dunlap Department of Astronomy \& Astrophysics \\ 
University of Toronto}

\author{Yansong Qian}
\affiliation{David A. Dunlap Department of Astronomy \& Astrophysics \\ 
University of Toronto}

\begin{abstract}
    {\it Kepler} planets (including super-Earths and sub-Neptunes, from 1 to 4 Earth radii) are likely formed before the gaseous proto-planetary disks have dissipated, as are the Jovian planets. If the metal content in these disks resembles that in the host stars, one might expect {\it Kepler} planets to occur more frequently, and to be more massive, around metal-rich stars. Contrary to these expectations, we find that the radii of { \it Kepler} planets (a proxy for mass) are independent of host metallicity. Previous claims that larger planets prefer more metal-rich stars can be adequately explained by the combined facts that more massive stars tend to host bigger planets,  and that more massive stars are also more metal rich in the {\it Kepler} sample. We interpret this independence as that the mass of a {\it Kepler} planet is not determined by  the availability of solids, but is instead regulated by an as yet unknown process. Moreover, we find that the occurrence rates of {\it Kepler} planets rise only weakly with stellar metallicity, a trend that is further flattened when the influence of close stellar binaries is accounted for. We explain this weak dependence, in contrast to the strong dependence exhibited by Jovian planets, using a phenomenological model, wherein the masses of proto-planetary disks have a much larger spread than the spread in stellar metallicity, and wherein the formation of Jovian planets requires disks that contain some $5$ times more solid than that needed to form {\it Kepler} planets.  This model predicts that stars more metal-poor than half-solar should rarely host any Kepler planets.
\end{abstract}


\section{Introduction}\label{sec:Intro}

The NASA {\it Kepler} mission found an abundance of small planets with radii smaller than $4 R_\oplus$ and orbital periods shorter than a year. These objects appear to occur around 30$\%$ of Sun-like stars \citep{Zhu2018a}.
According to their sizes, they have been further divided into the so-called super-Earths and mini-Neptunes sub-populations. 
Following a few previous works, we refer to them summarily as the ‘{\it Kepler} planets’. This is both for simplicity and for the following reasons. Even though they are not the entirety of {\it Kepler's} planet finds, they dominate in number. Moreover, these two groups likely are one and the same---it is argued that the super-Earths are simply the photo-evaporated remains of the sub-Neptunes \citep{Wu2013,Owen2013,Lopez2013,Jin2014,Chen2016,Owen2017}.

Absent from our own Solar System, the origins of these populous objects are currently unknown. However, since the sub-Neptunes (and presumably the young super-Earths) contain  sizable H/He envelopes, it is reasonable to assume that they were produced before the gaseous proto-planetary disks have fully dissipated. As such, these disks control the formation and orbital evolution of these planets. In fact, if we classify all planets into two classes, generation-I planets which are produced before the disk dispersal, and generation-II planets which form after, the \textit{Kepler} planets should belong to the gen-I group, together with Jovian giant planets, while terrestrial planets in the Solar System likely belong to the gen-II group. 

A natural way to study the formation of {\it Kepler} planets is to characterize how the properties of these planets relate to those of their host stars. In this article, we focus on one stellar property: stellar metallicity ($Z$). Planet properties 
can depend on $Z$ in two independent ways. The first one, studied extensively in the literature, is how planet occurrences depend on $Z$ (the occurrence-$Z$ relation).  The second is how planet sizes (and by proxy, masses) depend on $Z$. We call this the radius-$Z$ relation.  For testing models of planet formation, the latter is as powerful as the former. 

\subsection{Model Expectations}

Here, we list briefly theoretical expectations for these correlations, for a few representative formation models. Where these models fail to make explicit predictions, we infer them from the  descriptions in the original texts. 
We assume that a more metal-rich star has a more metal-rich disk and thus a higher initial solid content.

\begin{enumerate}

\item 
\citet{Hansen2012} suggested that {\it Kepler} planets are formed similarly as that proposed for terrestrial planets. Starting from a solid-rich disk, planetary embryos collide and coalesce until most of the mass has been incorporated into a few large cores, and a dynamically stable configuration has been reached. To reproduce the {\it Kepler} planets (each of mass $5-10 M_\oplus$) in the inner region where we observe them today, the solid density must exceed that of the minimum-mass-solar-nebula \citep{Weidenschilling1977,Hayashi1981} by at least an order of magnitude. This enhancement could either result from a higher metal content in planet-forming disks, or from radial migration of solids from outer regions. For the former, one naturally expects a positive occurrence-$Z$ relation. One also expects that larger cores be formed around more metal-rich stars.

\item 
\citet{Ida2004a,Ida2004b,Ida2005,Mordasini2009,Mordasini2012,Mordasini2015} presented population synthesis models for planet formation, where they inserted one planetary embryo at a few AU into each gaseous protoplanetary disk 
and followed the processes of core growth, envelope accretion and inward migration. In general, such models fail to predict the abundance of {\it Kepler} planets\citep{Ida2005}, although more recent models \citep{Mordasini2015}, including effects of torque saturation on type-I migration, have been more successful. 

With respect to the metallicity correlation, these works report that giant planets arise more frequently in metal-rich disks, because these disks can produce more massive cores at earlier times, allowing for runaway gas accretion. For Neptune-class objects, the so-called `failed cores', there appears to be little correlation between their presence and $Z$ \citep{Mordasini2012}. This is partly the result of their one-embryo policy: the single embryo can turn into a Jupiter in metal-rich disks, a Neptune in less rich disks, or a low-mass core in metal-poor systems. If multiple embryos were allowed, it is likely that metal-rich disks can host both Jovian and Neptune-like planets.  If this were the case, then the average metallicity of Neptune hosts would increase.
Furthermore, the higher metallicity may allow these Neptunes to grow to larger masses.

\item  
After the {\it Kepler} planets were discovered, much attention has  been given to `pebble' accretion to explain their formation \citep[e.g.][]{Lambrechts2014,Lambrechts2019,Chatterjee2014}. In these scenarios, small dust conglomerates (`pebbles') suffer strong aerodynamic drag and migrate inward rapidly. They can accrete efficiently onto planetary embryos (if they are present), or accumulate at a pressure bump near the star to form a new planet. When the planet mass reaches the so-called `pebble-isolation mass' \citep{Morbidelli2012,Lambrechts2014}, it can carve away enough gas from its vicinity to produce an exterior pressure bump which stalls further pebble accretion and initiates formation of the next planet. \citet{Lambrechts2019} argued that a higher pebble flux can more efficiently produce {\it Kepler} planets. Since this flux is likely associated with disk mass and disk metal-content, one expects a positive correlation between the occurrence of {\it Kepler} planets and $Z$, while little relation between their masses and $Z$.

\end{enumerate}

So, theories in general predict a positive occurrence-$Z$ relation, but differ in the radius-$Z$ prediction.  This motivates us to measure both correlations in the data.

\subsection{Twin Tests}

The occurrence-$Z$ correlation has long been established for giant planets \citep{Gonzalez1997}. These planets strongly prefer metal-rich stars with an occurrence rate that scales super-linearly with $Z$, possibly as $Z^2$ \citep[e.g.,][]{Santos2004, Santos2005,Fischer2005,Wang2015c,Petigura2018}. This scaling has been interpreted to support the core-accretion scenario, where large cores that can accrete gas to become giant planets should form more readily in solid-rich disks due to the abundance of raw material.

Giant planets are theorized to have solid cores with masses between $10-20 M_\oplus$ \citep[e.g.,][]{Pollack,Rafikov}, only a couple times heavier than the masses of {\it Kepler} planets \citep[$5-10 M_\oplus$, see e.g.][]{WeissMarcy,Wu2019}. This fact, together with the theories discussed above, lead one to expect a similarly strong occurrence-$Z$ correlation for {\it Kepler} planets. Transit and radial velocity studies to date, however,  suggest a weak or non-existent correlation \citep{Sousa2008,Sousa2011,Adibekyan2012,Buchhave2012,Buchhave2014,Buchhave2015,Wang2015c,Mulders2016,Sousa2018,Petigura2018,Zhu2019}. 

In this work, we further quantify the occurrence-$Z$  relationship. In particular, we account for the effects of close binaries, and construct a simple model to explain the different metallicity dependencies for {\it Kepler} and giant planets.  

In departure from earlier works that focus almost exclusively on the occurrence-$Z$ relation  \citep[for exceptions, see][]{Buchhave2012,Petigura2018}, here, we also investigate the radius-$Z$ relation. Unlike for giant planets, the sizes of {\it Kepler} planets can be readily used as proxies for their masses. While this is fairly obvious for super-Earths, its applicability for sub-Neptunes, which have hydrogen-rich atmospheres, is less obvious. Models of photo-evaporation have successfully reproduced the size distribution of {\it Kepler} planets. In these models \citep[e.g.][]{Owen2017}, the masses of sub-Neptunes are dominated by their cores, and their core radii are roughly a constant fraction (about half) of their total radii.\footnote{Our conclusions are not sensitive to the exact fraction, as long as it is independent of metallicity.} As such, the empirical radius-Z correlation can be translated to a planet mass-Z correlation, a direct constraint on planet formation.

When studying the radius-$Z$ relation, there is an interesting subtlety that needs to be considered.  Using the updated stellar parameters from the Gaia satellite, \citet{Fulton2018} first  inferred that the radii of Kepler planets rise with stellar mass. This is  confirmed by the analysis in \citet{Wu2019}, where she further argued that Kepler planets are formed tightly around a particular mass scale, the so-called `thermal mass'.  If planet masses are indeed correlated with stellar masses,  an issue we  further substantiate in this work, a secondary radius-$Z$ relation can arise: since more massive stars in the CKS sample appear to be more metal rich (a puzzle we address in Appendix  \ref{ap:massmet}), one would observe an apparent correlation where more massive planets tend to orbit more metal-rich stars, even if metallicity does not affect planet formation. There is a further twist. If, on the other hand, metallicity does matter and more metal-rich hosts tend to breed more massive planets, the above mentioned $M_*-Z$ correlation will yield an apparent correlation between stellar mass and planet mass. In fact, \citet{Fulton2018} cautioned against exactly this possibility.  

Which correlation is inherent and which is apparent? In this work, we aim to disentangle these effects to determine whether stellar mass, or stellar metallicity, is the main underlying cause for different planet sizes.

The layout of this paper is as follows: we first describe the sample for our work in \S \ref{sec:samp}; we then study the planet radius-$Z$ relation in \S \ref{sec:size}, and the occurrence-$Z$ relation in \S \ref{sec:occ}. Implications of our results are discussed in \S \ref{sec:discussion}. 

\section{Sample Selection}\label{sec:samp}

To perform the above tests, we require a sample of stars with known transiting planets, and a sample without. Both should have metallicity measurements, and they  should be as similar as possible in every way, except for the presence/absence of known transiting planets. We require the planet sample to have high purity and to have small error bars on planet radius, so as to tease out subtle relationships between the planets and their stars. These considerations lead us to the following choices.

The California Kepler Survey has measured metallicities for a magnitude-limited sample of planet hosts \citep{Petigura2017,Johnson2017,Petigura2018}. The metallicities reported by the CKS and LAMOST surveys are in terms of [Fe/H]. From these we infer stellar metallicity assuming [Fe/H]$=\log_{\rm 10}(Z/Z_\odot)$, where $Z_\odot$ is the solar metallicity.
These hosts are sun-like dwarf stars with Kepler magnitudes $K_p<14.2$. Planet candidates are those that obey the selection criteria in \citet[][\S 4.2]{Fulton2018},  excluding those with grazing transits or are false positives. Parameters for stellar mass and radius were later refined by \citet{Fulton2018} using {\it Gaia} parallaxes, reducing the error on planet radii to $\sim 5\%$. This is the so-called CKS-VII sample and it contains $907$ planet candidates. We adopt this sample as our planet sample. We further remove planets larger than $4R_\oplus$ and smaller than $1R_\oplus$, and are left with a total of  $696$ planet candidates orbiting $492$ stars. 

As part of the LAMOST \citep[Large Sky Area Multi-Object Fiber Spectroscopic Telescope---also called the Guo Shou Jing Telescope,][]{Zhao2012,Cui2012} DR4 \footnote{http://dr4.lamost.org/} survey, metallicities and stellar parameters for many stars in the {\it Kepler} field are also obtained. Before we adopt these as our non-transiting sample, we further remove any known planet hosts, and limit ourselves to FKG dwarfs with the same magnitude limit and the same effective temperature  range as for the CKS-VII sample. Furthermore, metallicity measurements from CKS and from LAMOST are known to have some offsets. We modify the LAMOST values based on the calibration performed in \citet[Appendix A of][]{Petigura2018}. This leaves us with a large sample of $21,962$ stars.

One important issue is the possible presence of selection effects. For instance, more massive stars have larger stellar radii,  potentially making it more difficult to detect small transiting planets. Detection completeness for the \textit{Kepler} pipeline has been well characterized by  \citet{Burke2017,Christiansen2020}. We address this issue in \S \ref{sec:completeness}. 

\section{Radius-Metallicity Relation}\label{sec:size}

Each {\it Kepler} planet in our  sample is quantified by four parameters: host star  mass, host star metallicity, planet radius, and orbital period. We employ a clustering analysis to investigate the relationships between all these parameters. We then focus on the effect of stellar metallicity.

\begin{figure*}
	\centering
    \begin{overpic}[width=0.85\textwidth]
        {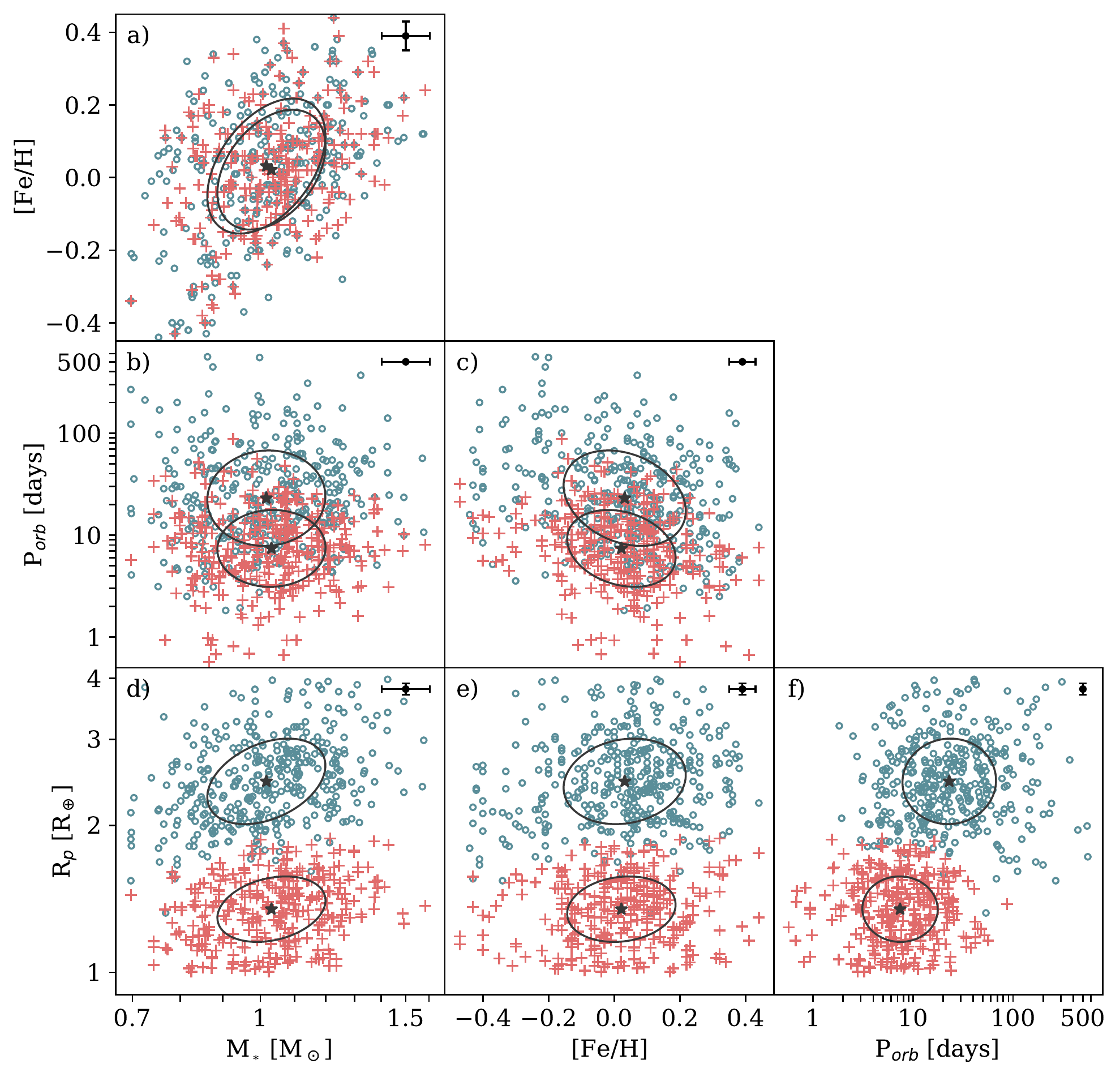}
        \put (30,82) { 
            \begin{tabular}{cc|cccccc}
                \multicolumn{8}{c}{Pearson's $\rho$ and associated p-values}\\
                \hline\hline 
                 &  & \multicolumn{2}{c}{Super-Earths} & \multicolumn{2}{c}{Sub-Neptunes} & \multicolumn{2}{c}{Combined} \\ 
               Pair & Panel & $\rho$ & $p$ & $\rho$ & $p$& $\rho$ & $p$\\
                \hline 
                 [Fe/H]-$M_*$ & a)  & 0.39  & $ 10^{-13}$ & 0.45  & $10^{-20}$ & $0.42$   &  ${\bf 10^{-32}}$ \\
               $P_{\rm orb}-M_*$ &  b)  & 0.02  & 0.66       & 0.04  & 0.45       & $ 0.38 $ & $0.03$ \\
               $P_{\rm orb}-$[Fe/H] & c)  & -0.24 & $ 10^{-6}$  & -0.27 & $10^{-7}$  & $ -0.26$ & ${\bf 10^{-12}} $\\
               $R_p - M_*$ &  d)  & 0.27  & $ 10^{-6}$  & 0.37  & $10^{-13}$ & $ 0.33$  &  ${\bf 10^{-19}}$\\
                $R_p-$[Fe/H] & e)  & 0.13  & $ 10^{-2}$  & 0.15  & $ 10^{-3}$  & $ 0.14$  & $ {\bf 10^{-4}}$\\
                $R_p - P_{\rm orb}$ & f)  & -0.01 & 0.82       & 0.02  & 0.66       & $0.01 $  & $0.79$\\
                \hline\hline
            \end{tabular}
        }
    \end{overpic}
    \caption{Correlation between planet and stellar properties, analysed using the {\it Mclust} package. Black error bars in the top right corners indicate typical measurement errors  as determined by \citet{Fulton2018}. Red crosses and green circles represent the super-Earth and sub-Neptune sub-populations, respectively, as identified by the clustering analysis. In each panel, the black ellipse graphically illustrates the covariance matrix within each population. The width of the ellipse represents dispersion in each parameter, while the slant the direction of maximum variation. The super-Earths and sub-Neptunes behave similarly in all panels.  The inset table shows the Pearson correlation coefficient ($\rho$) and $p$-value obtained for each sub-population, and for the combined population (after aligning the means) , for each pair of variables. We identify four statistically significant correlations (bold-faced in the table): panel a), more massive stars are also more metal rich;  panel c), planets orbiting metal rich stars have shorter periods; panel d), more massive stars host larger planets; panel e), a moderate correlation between planet radius and stellar metallicity.} 
    \label{fig:GMM}
\end{figure*}

\subsection{Clustering Analysis} \label{sec:clust}

Planet data are presented in Fig. \ref{fig:GMM}. All parameters are plotted in logarithms, since we assume the underlying relations are power-law in nature.

\begin{figure*}
    \centering
    \includegraphics[width=\textwidth]{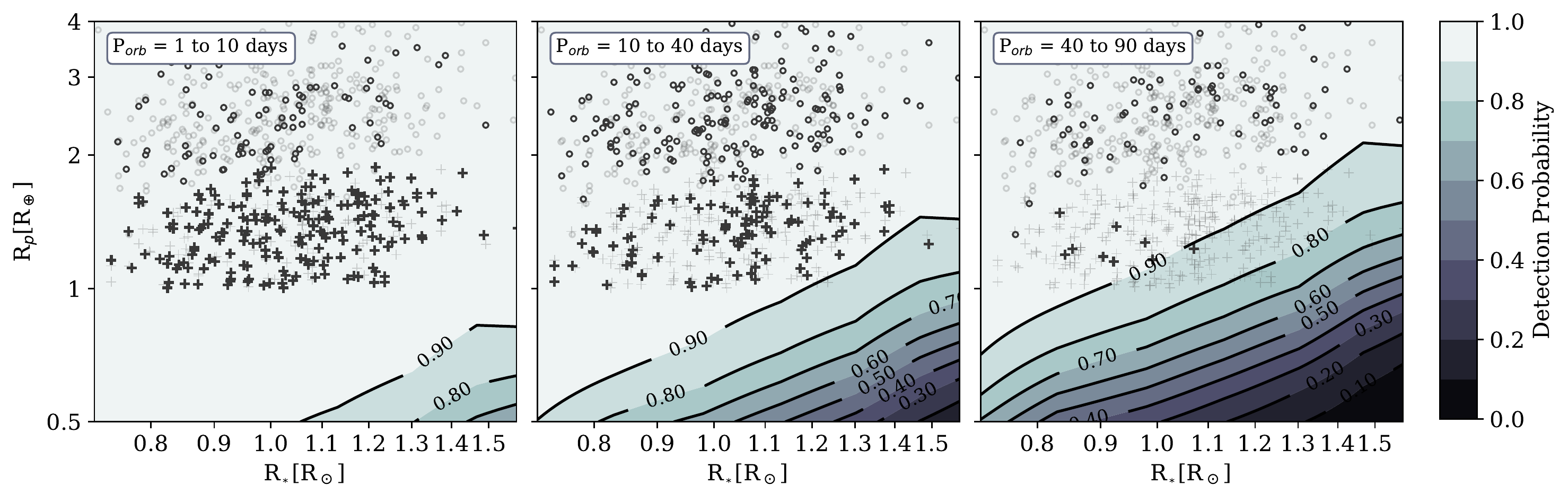}
    \caption{Sizes of the observed planets vs stellar radius, split into three  period ranges (bold points). Overplotted are the pipeline completeness contours. These are obtained by first calculating the completeness as a function of planet size and period for a given  star, then average these values over $400$ stars in each stellar radius bin, and lastly, average over the respective (logarithmic) period bins. The detection of super-Earths are only  mildly affected in the right-most panel, where a small smattering of super-Earths are found.
    }
    \label{fig:complt}
\end{figure*}
The data show clear substructures, reflecting the presence of sub-populations. To avoid the analysis be dominated by intra-group differences, we employ the \textit{Mclust} package  \citep[R-environment,][]{mclust,Fraley2002} to perform a clustering analysis. This package uses Gaussian Mixture Models (GMM) to classify objects into different populations, and obtain parameter correlations within any given population.  Guided by previous studies, we limit the total number of components to 3 in \textit{Mclust} to avoid over-fitting. The best model (with the highest Bayesian Information Criterion, or BIC) indeed requires two populations, ones that have been known as super-Earths and sub-Neptunes. They are roughly equal in numbers, with the super-Earths being smaller and closer to the stars, in agreement with results from many previous studies \citep[e.g.][]{Wu2013,Petigura2013,Fulton2017,Rogers2015}.

\textit{Mclust} also returns a variance matrix between variables $x$ and $y$, for each population, with the diagonal matrix elements ($\sigma_x^2, \sigma_y^2$, where $\sigma$ is the standard deviation) describing the variances in a given parameter, and the off-diagonal terms ($\sigma_{xy}$) the co-variances between two parameters. These are illustrated in Fig. \ref{fig:GMM} by the widths (diagonal) and the slants (off-diagonal) of the ellipses. The two populations, super-Earths and sub-Neptunes, are described by ellipses that are similar in shape and in orientation in all panels, confirming the hypothesis that they have the same origins and only differ in their later evolution (photo-evaporation of the atmospheres for the super-Earths). 

From the variance matrix, we obtain the Pearson correlation coefficients as 
\begin{equation} 
   \rho_{i,j}
   { = \frac{\sigma_{ij}}{\sigma_i\sigma_j} }\, .\label{eq:corr}
\end{equation} 
Since the Pearson correlation coefficient has the same assumptions as the GMM,  namely, linear relationships between the variables and Gaussian scatters, we use it over other correlation measures. In Fig. \ref{fig:GMM}, we also report the statistical significance ($p$-value) from a two-sided t-test for the null hypothesis that $\rho = 0$.\footnote{ P-values from Kendall's tau are in agreement with those reported in Fig. \ref{fig:GMM}.} We find that, for every pair of variables, the two sub-populations exhibit similar correlation properties. We report the $\rho$ and $p$-values for individual sub-population, and for the combined population. For the latter, we first remove the mean values for each cluster and then combine the sample.


This exercise uncovers the following four significant correlations, discussed in descending order of significance:
\begin{enumerate}
    \item 
    Masses and metallicities of the stellar hosts are significantly correlated, with a p-value of $p = 10^{-32}$. The trend can be described by a scaling of\footnote{To obtain this, we note that if $x$ and $y$ (logarithms of the observed quantities) both have zero means ($<x>=<y>=0$), and are linearly correlated, aside from a normally distributed scatter around zero ($y'$), i.e.,  $y=  \alpha x + y'$, then $\alpha = \sigma_{xy}/\sigma_{x}^2$.} 
    \begin{equation}
        Z \propto M_*^{1.08\pm 0.09} \, ,
        \label{eq:starz}
    \end{equation}
    with the power-law indices being  $0.99\pm0.19$ and $1.17\pm0.14$ for super-earths and sub-Neptunes, respectively.
    This trend is largely driven by the lack of high mass stars at low metallicities. However, given the narrow range of stellar masses in the sample, this is unlikely to be associated with any galactic evolution of metallicity. Instead, our analysis (detailed in Appendix \ref{ap:massmet}) ascribes this apparent correlation to sample selection in {\it Kepler} and CKS.
    
\item 
    Planet size correlates with host mass significantly, with $p = 10^{-19}$. Radius scales with mass as
    \begin{equation} 
        { R_{\rm p}\propto M_*^{0.29 \pm 0.09}}\, ,\\ 
        \label{eq:rplmass}
    \end{equation}
    with the power-law indices being $0.27\pm0.07$ for super-Earths and $0.45\pm0.10$ for sub-Neptunes.
    More massive stars appear to host larger (and therefore more massive) planets.  For a crude mass-radius relation of  $M_{\rm p} \propto R_{\rm p}^4$, this translates to $M_{\rm p} \propto M_*^{1.16 \pm 0.36}$, consistent with that found in \citet{Wu2019}. That paper further argued that this scaling is associated with the so-called `thermal mass' and is the direct result of planet formation. Reviewers of our paper have, however, worried that this may instead arise from observational selections. We address this carefully in \S  \ref{sec:completeness}.
    
    \item
    Another significant trend is found between metallicity and orbital period ($p =10^{-12}$), where planets around more metal-rich stars tend to orbit at shorter periods. This trend has been noted before by \citet{Mulders2016,Owen2018,Petigura2018,Dong2018} and remains unexplained.
    
    

\item Of particular interest to us is the modestly significant ($p=10^{-4}$) correlation between planet size and stellar metallicity. 
($p=10^-2$ and $p=10^{-3}$ 
for super-Earths and sub-Neptunes, respectively). Such a correlation has also been reported earlier  \citep{Buchhave2012,Buchhave2014,Petigura2018,Owen2018}. However, given correlation 1 and 2 above, it seems natural to wonder if this correlation is merely a  by-product of those two, without reflecting the intrinsic planet formation. This is the topic we address in  \S  \ref{sec:mock}.   
    \end{enumerate}

\subsection{Detection Completeness}
\label{sec:completeness}

Could the trend that planets around more massive stars tend to be larger be caused by selection bias? More massive stars are bigger and therefore it may be  harder to detect transits of smaller planets. Here, we compute realistic transit detection completeness following the procedure in \citet{QianWu}, to demonstrate that the observed trend is genuine.\footnote{This analysis supersedes an earlier attempt by \citet{Wu2019} that was  based on an incomplete hand-waving argument.} 

We use the \textsc{KeplerPorts} software \citep{Burke2017} to compute the probability of pipeline detection when a planet with a certain size and period transit in front of a star. This detection efficiency model is obtained through signal injections on the observed light-curves (not pixel-level), it can therefore account for stellar noise and target-to-target variations.
We adopt the stellar sample as characterized by \citet{Berger2020}, subject it to the same cuts as for the CKS sample \citep[as described in Table 2 of][ for FGK dwarfs]{Petigura2018}. We then split stars into 14 radius bins from  $0.6$ to $1.6 R_\odot$, roughly  corresponding to mass bins of the same values. To determine the characteristic detection efficiencies for stars in each bin, we average values over $400$ randomly drawn stars. These efficiencies are used in the following two exercises to demonstrate our thesis.

First, we overplot  our planet sample with contours of detection efficiency  
(Fig. \ref{fig:complt}), in the space spanned by planet radius and stellar radius. To illustrate the effects of period, we split planets into three groups ($1-10$ days, $10-40$ days and $40-90$days). About $60\%$,
$97\%$ and $99.9\%$ of the observed super-Earths are contained within $10, 40$ and $90$ days, respectively  (Fig.\ref{fig:GMM}).
The completeness curves, obtained by linearly averaging over these respective period bins, show that even super-Earths, the smaller of the two populations, suffer little incompleteness. 
In the two shorter period bins, all observed super-Earths lie above the $90\%$ completeness curve, while in the longest period bin, super-Earths are harder to detect around the largest stars, but even there, the completeness hovers around $80-90\%$ and the overall effect is minor because few  super-Earths have such long periods. The detectability of sub-Neptunes is nearly perfect across all bins.

\begin{figure}
    \centering
    \includegraphics[width=0.45\textwidth]{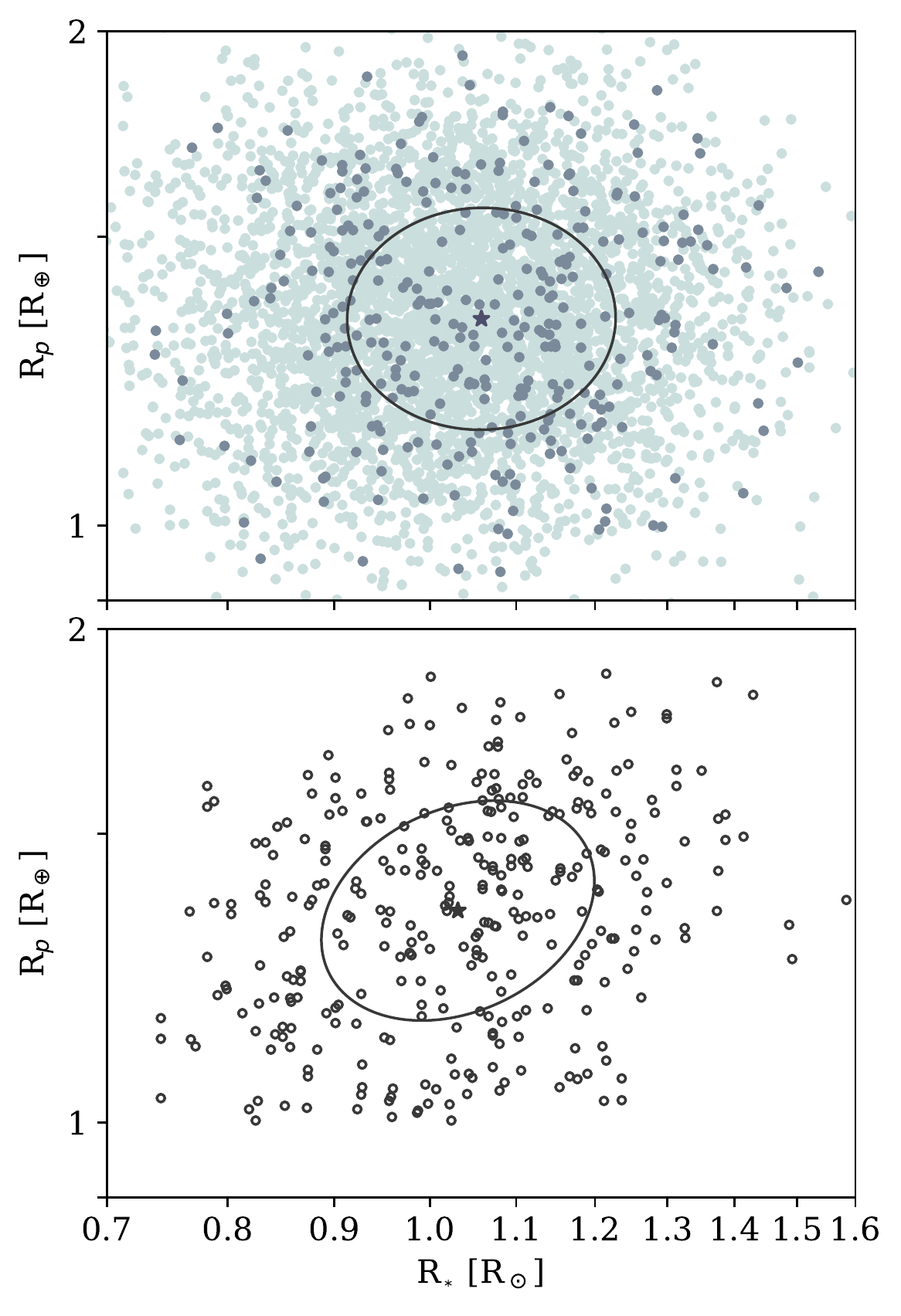}
    \caption{Mock (top panel: dark points are detected, light points are total population) and real (bottom panel) super-Earth samples, in planet radius vs. stellar radius plane. The mock sample is designed to mimic the real sample in every sense, except that the sizes of the super-Earths are constant across all  stars. Even after completeness correction, which depends on stellar properties, are applied, it exhibits no detectable correlation. This in contrast to the observed one.
    \label{fig:sim_sample}}
\end{figure}

We use a second test to demonstrate that the observed $R_p - M_*$ correlation is not due to incompleteness. Let us assume that planets around all stars have the same size distribution, which is uncorrelated with stellar mass. We then  use the  fact that super-Earths ($\sim1.7 > R_p > 1 R_\oplus$) can be detected completely around smaller stars ($R_* < 1 R_\odot$) out to  at least $100$ days, to construct a planet sample. We demand that such a mock sample, after passing through a `transit pipeline' based on our pre-calculated transit probability and detection completeness, exhibit the same period distribution, the same stellar radius distribution, and the same planet number as the observed one. The radius distribution of such  a mock sample is shown in Fig. \ref{fig:sim_sample}. It shows no correlation between  planet size and stellar size, with a Pearson coefficient $\rho = 0.01$ and $p=0.5$, averaged over $200$ realizations. 

The third way to convince a sceptic that detection incompleteness does not cause the observed $R_p - M_*$ relation is to point to the sub-Neptunes. These planets are more completely detected than the  super-Earths are, yet they exhibit a similar (even a tad more significant) correlation.

\begin{figure*}
    \centering
    \includegraphics[width=\textwidth]{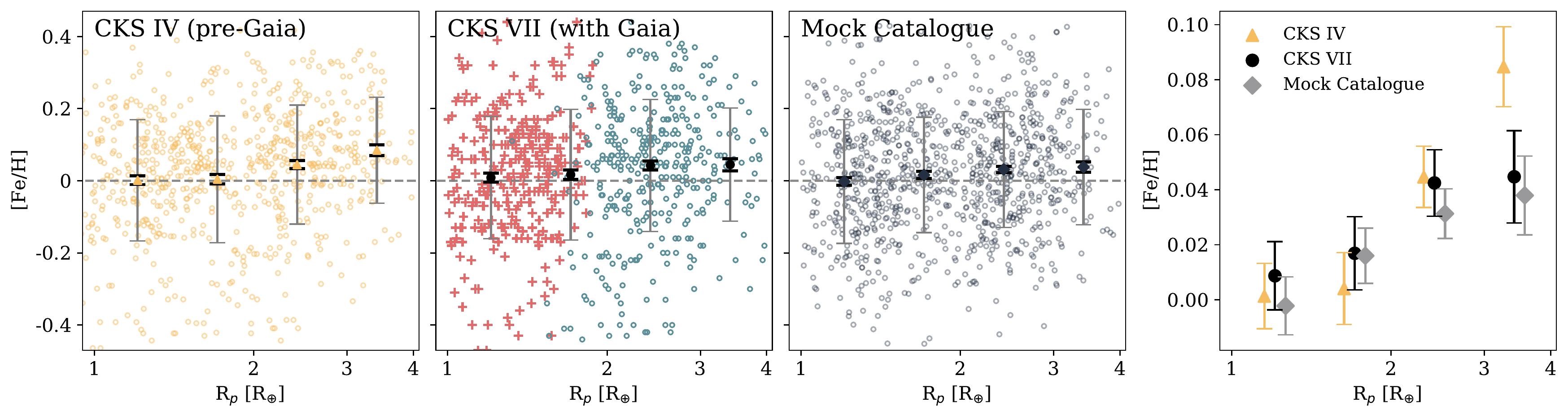}
    \caption{Further analysis of the radius-$Z$ relation. The three left panels show results from CKS VI \citep[pre-GAIA, Fig. 3 of][]{Petigura2018}, from CKS VII \citep[post-GAIA,][ used in this study]{Fulton2018}, and from our mock catalogue (see main text).
    Mean metallicities in each size bin are shown as filled symbols, with thin error bars indicating the standard deviations, and  thick error bars the standard errors.  These are again plotted in the right-most panel, displaced slightly horizontally for clarity of comparison. 
    While the pre-GAIA data shows a prominent rise in mean metallicities with planet sizes, such a trend is much blunted in the post-GAIA data. 
Moreover, the mock catalogue largely reproduces the blunted rise, demonstrating that the latter is merely a by-product of the planet size-stellar mass correlation. 
  }

    \label{fig:fakemetcorr}
\end{figure*}

\subsection{The Underlying Variable is Stellar Mass, not Metallicity }\label{sec:mock}

After ascertaining that planets are intrinsically larger around more massive stars, we return to consider whether this can explain the moderately significant correlation observed between planet size and metallicity (Fig. \ref{fig:GMM}), given that more massive stars in this sample are also more metal rich (see Appendix \ref{ap:massmet} for the origin of this correlation). 

A number of previous studies have shown that hosts of larger planets tend to be more metal rich \citep{Buchhave2012,Adibekyan2013,Buchhave2014,Petigura2018,Owen2018}. However, GAIA data brought about a change. As Fig. \ref{fig:fakemetcorr} shows,  post-GAIA data \citep{Fulton2018}, when compared with pre-GAIA data \citep{Petigura2018}, show a much 
milder rise in mean metallicity with planet size. Similarly, while 
\citet{Owen2018} adopted CKS I-III stellar parameters and found that the radius distributions of metal-rich and metal-poor planets differ with a significance of $p=10^{-5}$, the same exercise, using the updated parameters, now returns $p=0.07$.
In the following discussion, we show that even such a mild dependence on  metallicity can be explained away.

We produce a mock catalogue of planets that contain the same number of planets as observed. Each mock planet is assigned to a mock star with its properties (mass and metallicity) randomly sampled from those in panel a) of Fig. \ref{fig:GMM}, such that more massive stars are more metal-rich. As for the planet, we first assign it to a sub-population (super-Earth or mini-Neptune) based on the observed mixing coefficient, we then endow it with a size  such that we can reproduce the observed variance matrix between planet size and stellar mass. These steps preserve the observed correlations (eq. \ref{eq:starz} and \ref{eq:rplmass}), but the planet does not know about stellar metallicity except through stellar mass. The resulting mock sample reproduces the observed mild rise of mean metallicity with planet size (Fig. \ref{fig:fakemetcorr}). Performing a GMM analysis on the mock catalogue returns a Pearson coefficients of $\rho = 0.11$ and $p=10^{-4}$, nearly identical to those for the real sample. 

An additional support comes from the following test, prompted by one of our reviewers. We repeat the entire GMM analysis, but limiting ourselves to a narrower metallicity range of [Fe/H]$= [-0.3,0.3]$, We find that the combined p-value for the [Fe/H]-$M_*$ pair drops to $10^{-12}$, from the original  $10^{-32}$. Because we argue that this correlation (which itself is due to selection bias, Appendix \ref{ap:massmet}) drives the observed $R_p-M*$ correlation, we expect the latter to weaken substantially. This is indeed observed. The combined p-value is now insignificant, $p=0.18$. In contrast, the $R_p-Z$ pair remains strongly correlated ($p=10^{-14}$) and with a similar slant, suggesting that it is inherent and robust.

In summary, the radii of {\it Kepler} planets are determined by stellar mass, and are not influenced by stellar metallicity. The apparent  correlation between planet radii and stellar metallicity is entirely a by-product of the latter's correlation with stellar mass, which is itself a result of selection bias.

\section{The Occurrence-$Z$ Relation} \label{sec:occ}

\begin{figure}
    \centering
    \includegraphics[width=0.45\textwidth]{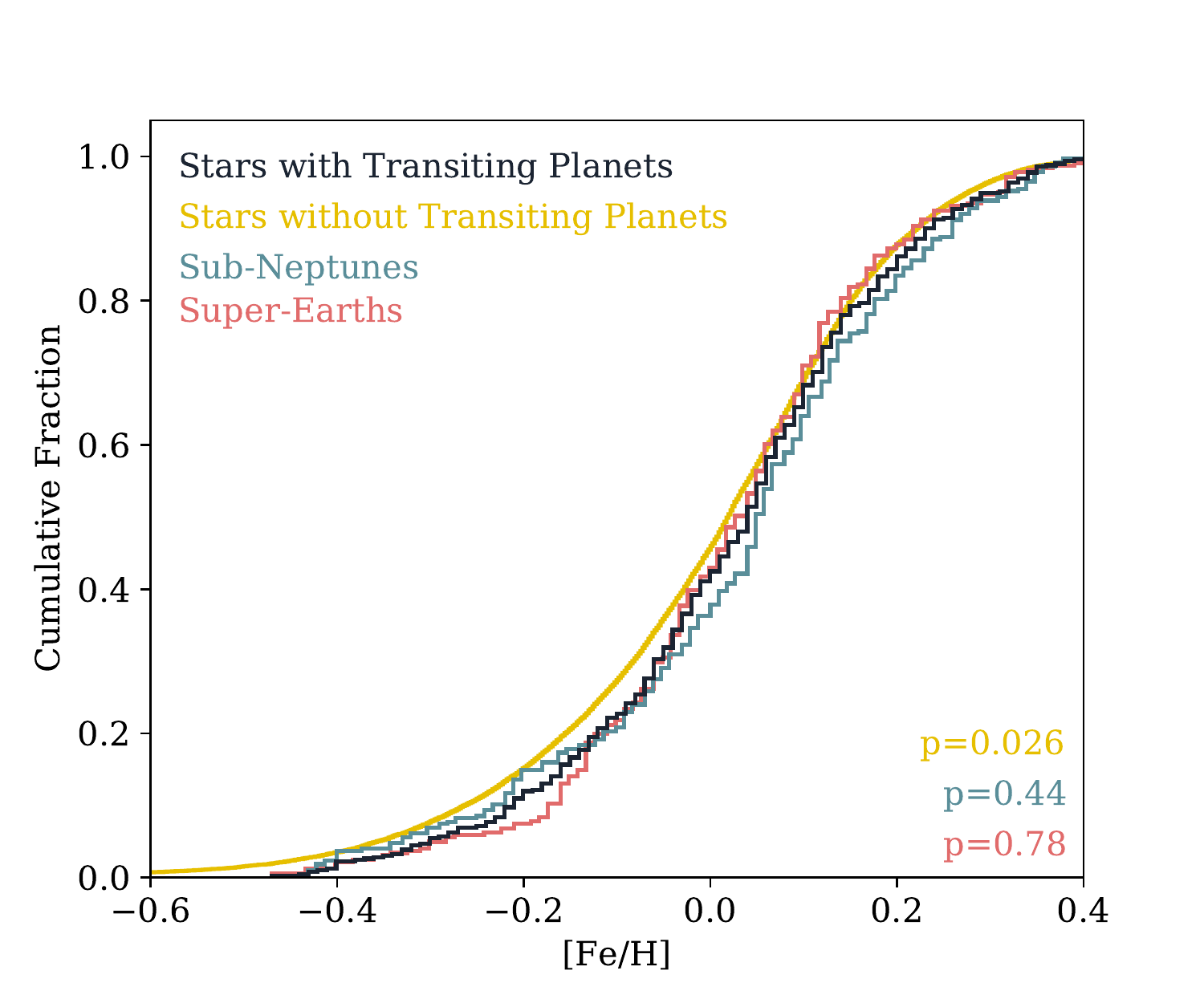}
    \caption{Comparing the cumulative metallicity distribution of stars with (black line) and without (yellow line) transiting planets. The former, comprising of $492$ planet hosts, are more metal-rich and are statistically different from the latter with a p-value of $0.026$. In contrast, hosts of sub-Neptunes and super-Earths are statistically indistinguishable from each other (p-values marked comparison with the black line). Stars that host both types of planets are counted once in each group.} 
    \label{fig:CDFocc}
\end{figure}

We now turn to the question of whether {\it Kepler} planets are preferentially formed around more metal-rich stars, the occurrence-$Z$ relation. Many previous studies have failed to take into account a number of complicating factors, warranting a new visit. The factors that may potentially influence our conclusion are:
\begin{enumerate}
    \item 
    It has been suggested that multiplicities in planetary systems depend on stellar metallicity \citep{Zhu2019}. To remove this effect, we count each star only once, regardless of the number of transiting planets it has. 

    \item 
    More massive stars (which tend to be more metal rich) are larger and it may be harder to detect transiting planets around them due to the shallower transit depth. This selection effect, if present, may skew the planet hosts toward the more metal poor side. Our discussions in \S \ref{sec:completeness} reject this bias.

    \item 
    For a given orbital period, planets around a more massive star has a smaller semi-major axis. This increases the transit probability for these planets. To account for this, we implement a forward model below.

    \item 
    The sample of stars that have no transiting planets may also contain planets (that are not transiting). This needs to be modelled (see below).

\end{enumerate}

In Fig. \ref{fig:CDFocc}, we compare the { $492$} planet hosts in the CKS sample to the 21,962 field stars that have no known transiting planets. Their cumulative metallicity distributions are statistically different ($p=0.026$), in agreement with previous results \citep[see, e.g.,][]{Zhu2019}\footnote{This comparison is subject to systematic differences between metallicity derivations from CKS and LAMOST. We account for these by following Appendix 1 of \citet{Petigura2017}, but  some additional systematics may remain. }.  In contrast, host stars for super-Earths and for sub-Neptunes are statistically indistinguishable from each other, again supporting the suggestion that these two sub-populations share the same origins and only diverge in their late evolution. 

\subsection{Quantifying the Occurrence-$Z$ relation}

To better quantify the occurrence-$Z$ relation, and to properly account for the selection effects discussed above, we implement a forward model, largely following the strategy set out in  \citet{Zhu2016}, but with a few minor changes. 

As in \citet{Fischer2005}, we describe the fraction of stars with at least one {\it Kepler} planet by a power-law,
\begin{equation} 
    f(Z) = \alpha \left(\frac{Z}{Z_\odot}\right)^\gamma \, . 
    \label{eq:fz}
\end{equation}
Determining the index, $\gamma$, is the goal of this section.

The normalization factor $\alpha$ is obtained from the overall planet occurrence rate $\eta$ by
\begin{equation} 
    \eta = \int f(Z)\, g([{\rm Fe/H}])\, d [{\rm Fe/H}] \, ,
\end{equation} 
where $[{\rm Fe/H}] = \log(Z/Z_\odot)$, and $g([{\rm Fe/H]})$ is the distribution of stellar metallicity in the \textit{Kepler} field. LAMOST reports that $g([{\rm Fe/H}])$ can be roughly described by a log-normal function with a mean of $-0.03$ and a standard deviation of $0.2$. For $\eta$, we adopt the recent determination by \citet{Zhu2018a} of $\eta=0.3\pm0.03$. This value is smaller than previous estimates, so $f(Z)$ always falls below unity and there is no need to include a saturation metallicity as was done in \citet{Zhu2016}. 

We first produce a large number of mock stars according to the above metallicity distribution. Each star is then assigned a mass according to   the covariance properties of $M_*-Z$ (i.e. panel a of Fig \ref{fig:GMM}),
and a radius assuming the main-sequence mass-radius relation,
$R_* = R_\odot (M_*/M_\odot)$. This differs from that in  \citet{Zhu2016} where all stars are assumed to be sun-like. Based on eq. (\ref{eq:fz}), we further assign a planet-hosting status to the star, and then proceed to determine, in the case of a planet host, if its planet should be transiting or not. We let the orbital inclination be randomly distributed in $\cos i$ from $[-1, 1]$, and we assume an orbital period that is uniformly distributed in the logarithmic space, for  periods from 5 to 400 days. Transiting planets are those that satisfy  $\cos i <R_*/a$, where  $a$ is the semi-major axis. To imitate the observational uncertainties in CKS and LAMOST, we add a random Gaussian error to the stellar metallicity with dispersions of either ${\sigma}_{\rm [Fe/H]}=0.09$ (the LAMOST error) or  ${\sigma}_{\rm [Fe/H]}=0.04$ (CKS error).

\begin{figure}
    \centering
    \includegraphics[width=0.45\textwidth]{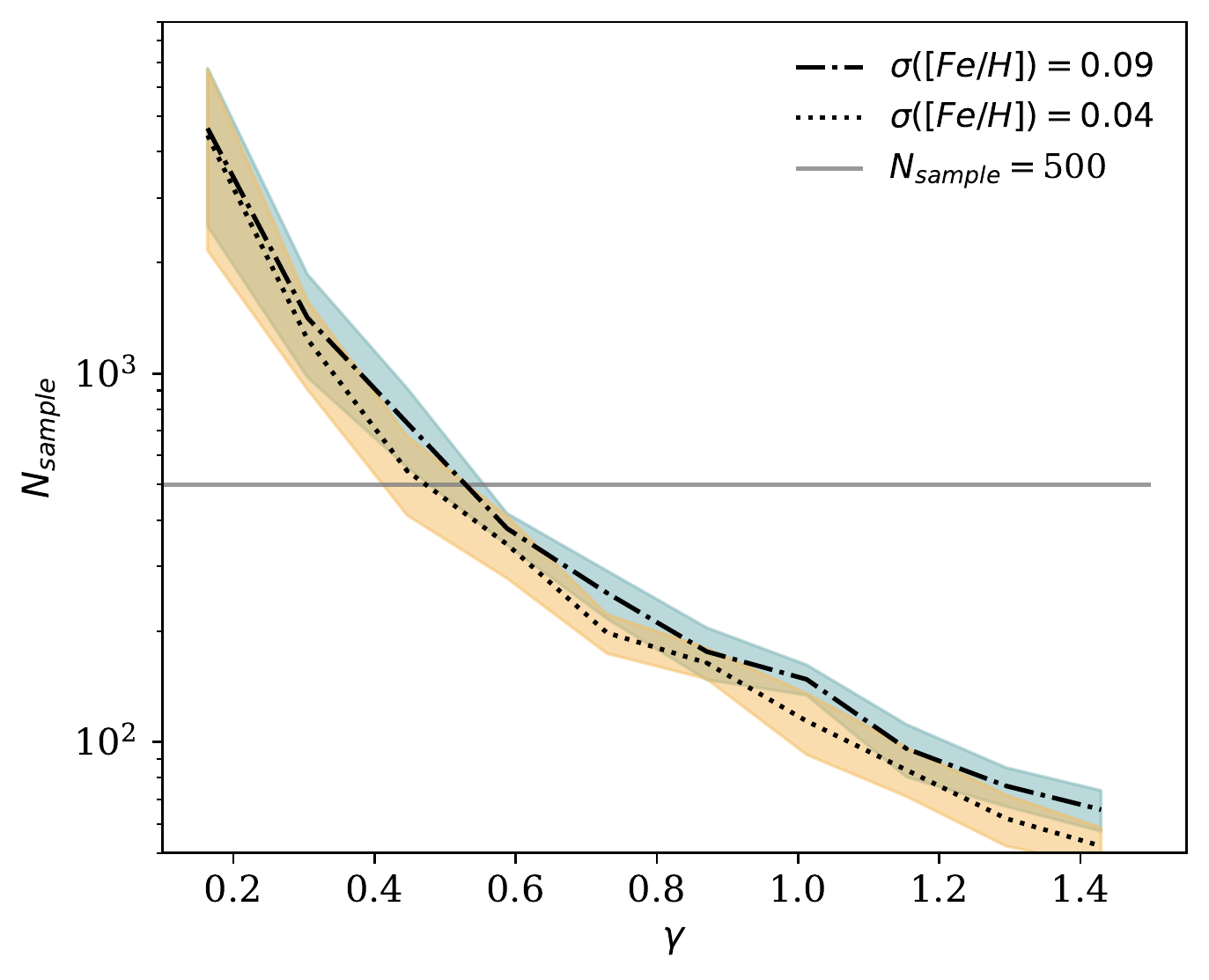}
    \caption{
    Quantifying the occurrence-$Z$ relation. The dotted and dashed lines indicate the sample size required to be able to distinguish transiting planet hosts from non-transiting stars, with a statistical significance that is comparable to the observed one ($p=0.026$), for different values of $\gamma$.  Shaded areas show the corresponding $1-\sigma$ spread due to uncertainties in metallicity measurements (LAMOST in blue and CKS in orange). 
    Since the observed planet sample size $N_{\rm sample} \sim 600$ (the solid black line), this yields $\gamma=0.55\pm0.1$, i.e., a sub-linear dependence between occurrence and metallicity.}
    \label{fig:fwrd}
\end{figure}

At each $\gamma$ value, we draw 200 random samples of transiting hosts (each of size $N_{\rm sample}$), and compare their metallicity distributions against the non-transiting sample. When $95\%$ these have p-values falling below our observed one  ($p=0.026$), we record the value of $N_{\rm sample}$. These are shown in Fig. \ref{fig:fwrd}. We find that for a sample size of  500 planet hosts (the CKS-VII sample), we can detect a metallicity dependence of the observed significance,  if $\gamma \approx 0.55\pm0.1$ and the value of $\alpha$ is $\alpha = 0.3\pm0.03$. 
Previously, \citet{Petigura2018} conducted a similar analysis for hot and warm super-Earths (with periods $P<10$, $10<P<100$ days respectively). But instead of using the fraction of stars that host planetary systems ($f(Z)$), they used the average number of planets per star as a proxy for the occurrence rate. This may affect the conclusion, if the number of planets per star depends on metallicity \citep{Zhu2019}. Nonetheless, their estimate of $\gamma = 0.6\pm0.2$ and $0.3\pm0.2$, for hot and warm super-Earths respectively, are similar to our value here. In a separate work, \cite{Zhu2019} adopted a different methodology to calculate the fraction of planetary systems as a function of host metallicity. He did not explicitly report the value of $\gamma$, but we estimate $\gamma\sim0.4$ from his Fig. 4, again consistent with our result. 

So the occurrence rate for {\it Kepler} planets scales sub-linearly ($\gamma = 0.55 \pm 0.1$) with stellar metallicity. 

\subsection{Effects of Close Binaries}\label{sec:plbin}

\begin{figure}
    \centering
    \includegraphics[width=0.45\textwidth]{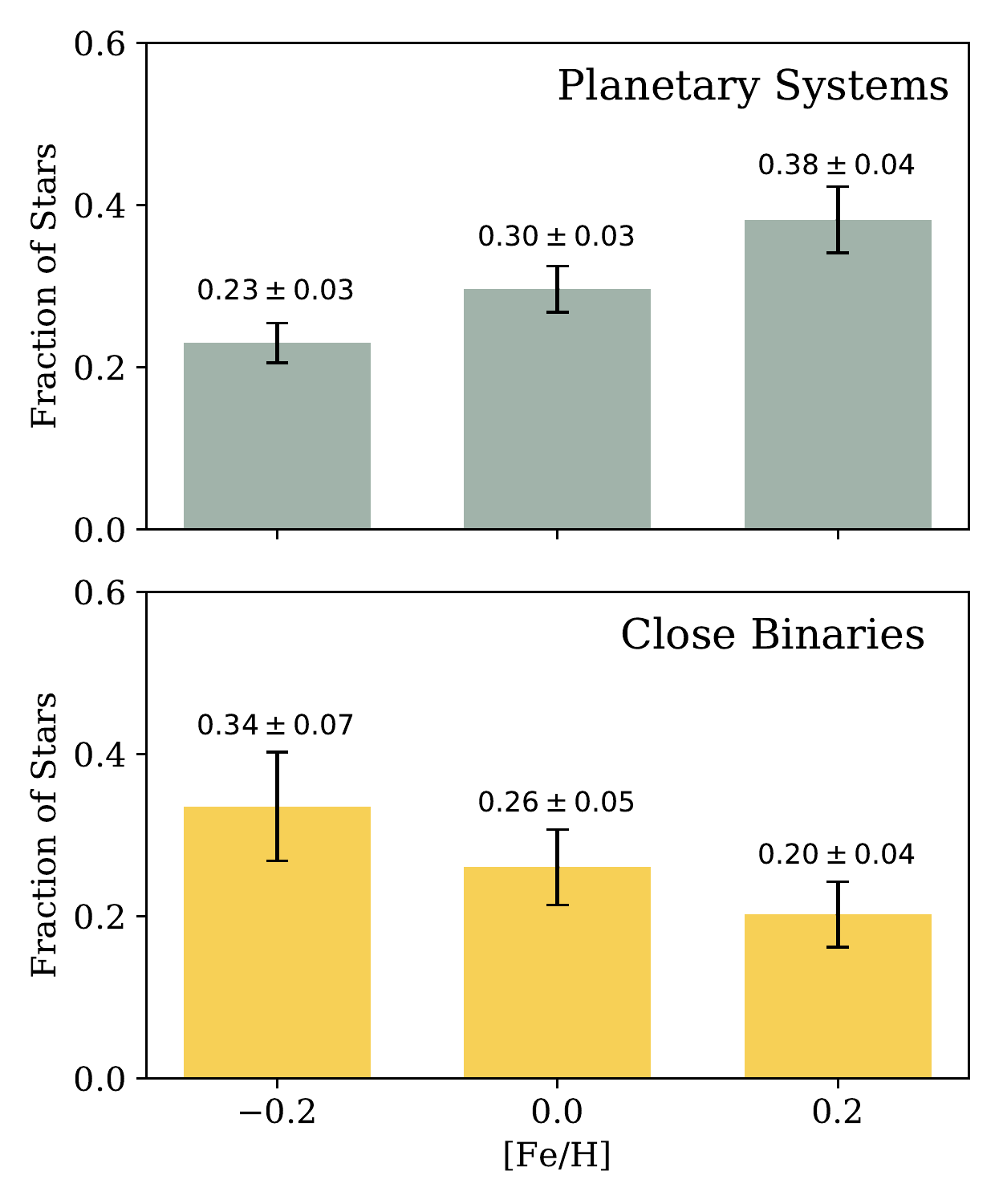}
    \caption{Fractions of stars that host {\it Kepler} planets (top panel) and close binaries (bottom), as functions of stellar metallicity. Close binaries here refer to those that can effectively suppress the formation of planets (ranging from 10-100AU in separations, see text). As these binaries prefer low metallicities, stars that can potentially form planets tend to be more metal-rich than average. This partly explains the positive metallicity trend observed for planets. 
   }
    \label{fig:binFrac}
\end{figure}

Here, we argue that close binaries may be partially responsible for the metallicity trend reported above.

The fraction of close-binaries ($P < 10^4$ days, or $a \leq 10-50$ AU) among sun-like stars appears to be strongly anti-correlated with stellar metallicity \citep[see review by][]{Moe2019,PriceWhelan2020,ElBadry2018},  with the fraction dropping from $24\%$ to $10\%$ when metallicity rises from $[Fe/H]=-0.2$ to $0.5$. This trend is thought to be related to the propensity for gravitational fragmentation in more metal poor disks (lower cooling time). 

At the same time, {\it Kepler} planet hosts are known to avoid close binaries. Adaptive optics imaging survey of these stars \citep{Wang2014,Armstrong2014,Wang2015a, Wang2015b,Kraus2016,Furlan2017,Ziegler2018,Matson2018} reported a striking paucity of close binaries when compared to field stars, while little difference in the fraction of wide binaries ($a > 50$AU). Interestingly, in the pre-main-sequence phase, close binaries are also observed to have an anomalously low disk fraction \citep{Kraus2012} when compared to single stars,  whereas wide binaries do not. So it appears that close binaries can disrupt their proto-planetary disks very early on in life \citep{Cieza2009,Kraus2012,Cheetham2015,Barenfeld2019}, thereby preventing the formation of {\it Kepler} planets . 

This effect alone can lead to a positive occurrence-$Z$ relation in the {\it Kepler} planets. To remove the impact of close binaries, we instead measure the occurrence as  
\begin{equation}
    f(Z)' = \frac{f(Z)}{1 - f_{\rm CB}(Z)} = {\alpha'} \left({\frac{Z}{Z_\odot}}\right)^{\gamma'} ,
\label{eq:fz2}
\end{equation}
where primed symbols denote those corrected for close-binary suppression, and $f_{\rm CB}(Z)$ is the fraction of stars that are in close binaries where planet formation is suppressed. 

For the three metallicity bins in Fig. \ref{fig:binFrac}, $f(Z)$ rises from $\sim 23\%$ to $\sim 38\%$ (adopting $\gamma= 0.55\pm 0.1$, and an overall planet fraction of $30\pm 3\%$). Meanwhile, we obtain the fraction of close binaries that can suppress planet formation from \citet{Moe2019a}. Binaries closer than $10$ AU can suppress planet formation completely, they have fractions $24 \pm 3\%$,  $19\pm 4\%$, $15\pm 3\%$, respectively, for the same metallicity bins. \citet{Moe2019a} further showed that binaries from $10$ to $100$ AU can have a gradated suppression, boosting $f_{\rm CB}$ by a factor of $1.4$ over the above cited fractions (see Fig. \ref{fig:binFrac}). Inserting these values into eq. (\ref{eq:fz2}) yields $\gamma' = 0.35 \pm 0.21$, a significant reduction.\footnote{ The value of $\gamma$ may be further reduced to $0.26\pm0.26$ if one includes the Malmquist bias, as argued in \citet{Moe2019a}.}

So while our original result shows that planet occurrence depends on $Z$ weakly (sub-linearly), this discussion argues that part of the dependence arises from planet suppression by close binaries that are preferrentially metal-poor. Among stars that can potentially form planets, metallicity plays a minor role. 

\subsection{A Crude Model}

\begin{figure*}
    \centering
    \includegraphics[width=0.45\textwidth,trim=20 140 50 115,clip]{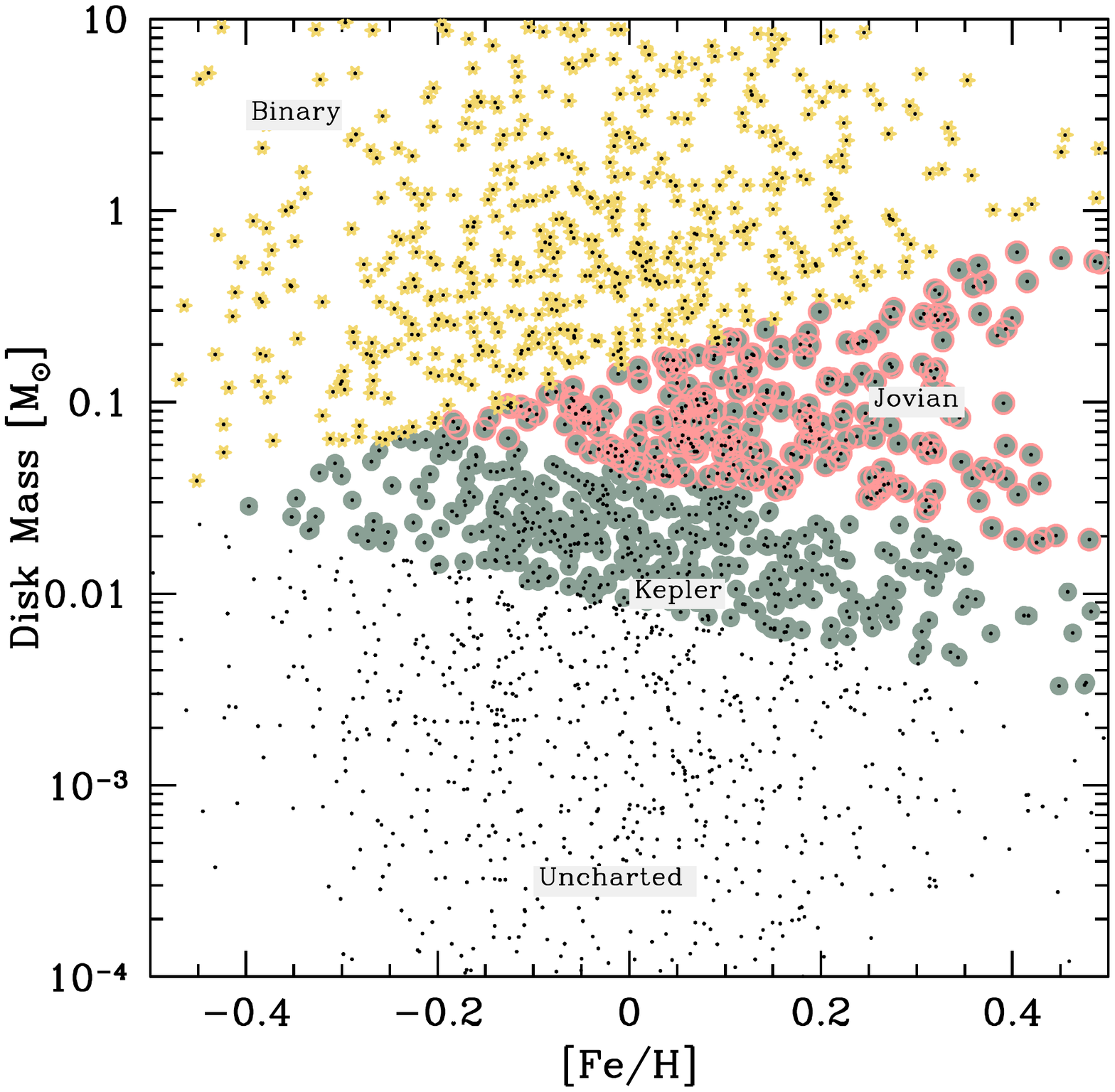}
        \includegraphics[width=0.45\textwidth,trim=20 140 50 115,clip]{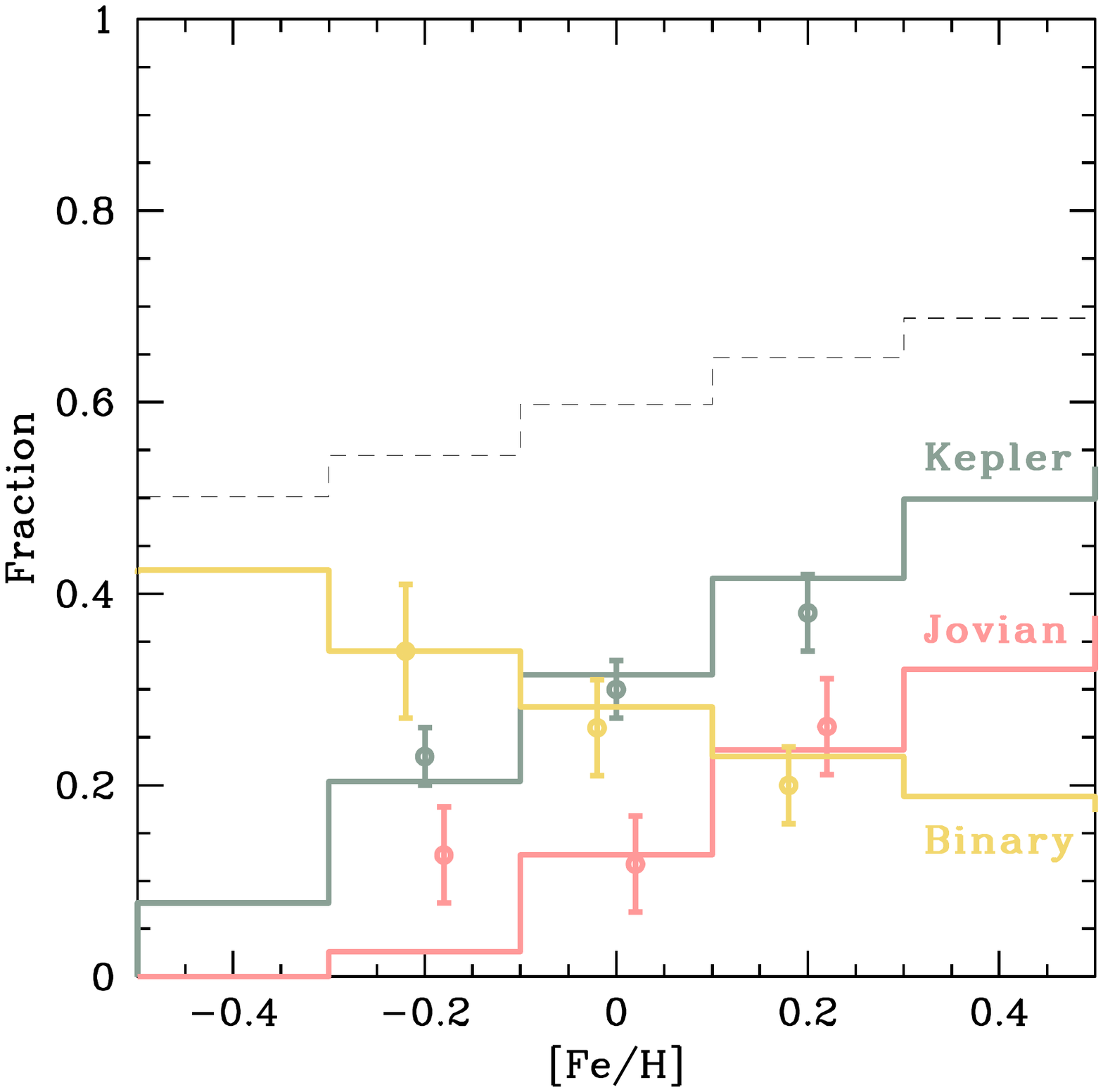}
    \caption{ A simple formation model that explains the occurrence-$Z$ relations for close binaries, {\it Kepler} planets and giant planets. The left panel illustrates how these populations are distributed in the disk-mass -- metallicity plane. Here, `Binary' refers to close binaries that can suppress planet formation, `Jovian' planets refer to giant planets outside $0.5$AU. They are found exclusively in systems that also form `Kepler' planets (super-Earths and sub-Neptunes). Right figure: the fractions of stars that host various populations, as functions of stellar metallicity. Model results are shown as solid histograms, while observations are marked as circles. 
    The dotted line is the combined fractions for close binaries and Kepler planets. Our simple model reproduces most data, except for the most metal-poor bin for Jovian planets.}
    \label{fig:scat}
\end{figure*}

Here, we construct a simple phenomenological model to explain the observed occurrence-$Z$ relation for {\it Kepler} planets. This model also accommodates the metallicity dependencies for giant planets and close binaries. 

As discussed in \S \ref{sec:Intro}, the core-accretion scenario predicts the core masses for giant planets, as required for run-away gas accretion, are $\sim 10-20M_\oplus$, while the masses of {\it Kepler} planets are about half of that. This is not a large difference. However, the fact that they differ markedly in their occurrence-$Z$ relation is surprising. One possible solution is to argue that solids in the protoplanetary disks form {\it Kepler} planets first, and only those that have an excess of material can go on to form giant planets. Studies by \citet{Zhu2018,Bryan2019,Herman2019} support this suggestion, where they found that giant planets appear to orbit mostly stars that also host inner Kepler planets. 

Moreover, if the total masses of protoplanetary disks have a large spread, much larger than that in metallicity, the dispersion in disk solids will be dominated by the former variation. This will then dilute any dependence planet formation may have on metallicity. 

Here, we first obtain the occurrences for the three populations of concern. Data for close binaries and for Kepler planets are as shown in Fig. \ref{fig:binFrac}. For the giant planets, we restrict ourselves to cold giant planets, since the hot and warm varieties are much rarer, they are also suspected to have elevated metallicity trends and may not be  representative  \citep{Dawson2013,Buchhave2018}. We query the exoplanet database for planets with minimum mass above $0.15 M_J$ and with orbital separation larger than $0.5$ AU.  These are predominantly discovered by the radial velocity technique, orbiting around FGK hosts in the solar neighbourhood. For the three metallicity bins as in Fig. \ref{fig:binFrac}, this yields $68, 98$ and $134$ objects, respectively. To calculate the fractional occurrences, we assume that the FGK stars in the solar neighbourhood have the same metallicity distribution as that in the Kepler field, namely, a Gaussian distribution in $[Fe/H]$ with a dispersion of $0.2$ and a mean of $-0.004$, after calibration to the CKS metallicity values \citep{Zhao2012,Cui2012}. To normalize, we adopt an average Jovian occurrence rate of $13\pm 5\%$, consistent with results from  \citet{Cumming2008}. The occurrence rates for the three metallicity bins are then $13\pm 5\%$, $12\pm 5\%$ and $26\pm 5\%$ (Fig. \ref{fig:scat}).

The following are details of our crude model. We assume a single  population of proto-planetary disks, with disk masses satisfying a normal distribution,
\begin{equation}
    {\frac{dN}{d\log M_{\rm disk}}} \propto \exp{\left[- \frac{(\log M_{\rm disk} - \log M_0)^2}{2 \sigma_{\log M}^2}\right]}
\, ,    \label{eq:diskmass}
\end{equation}
where the mean $M_0$ and the dispersion $\sigma_{\log M}$ are taken to be $M_0 = 0.02 M_\odot$ and $\sigma_{\log M}= 1.5$. These values are compatible with the median mass and mass spread observed for real disks
 \citep{Andrews2013,Mohanty2013,Ansdell2016}. As for all parameters listed below, these values are not meant to be best-fits (hence no error bars), but are crude estimates to roughly reproduce the observed data. They are only meant to be illustrative. 
 
 Let the total solid mass contained in each disk be \begin{equation}
M_{\rm solid} = 1\% \times \, M_{\rm disk} \times \left(\frac{Z}{Z_\odot}\right) \, ,
\label{eq:solidmass}
\end{equation}
where we implicitly assumes that the disk metallicity resembles the stellar metallicity. For the latter, we adopt the same as that for the {\it Kepler} field (see above). Our results on the occurrence-$Z$ relation are not affected by the actual $Z$ distribution that we adopt. 

The following ``birth conditions" are the most important assumptions in our model. 
\begin{itemize}
    \item Close binaries\footnote{Here, this refers to only those that can effectively suppress planet formation} are formed if 
    \begin{equation}
    M_{\rm disk} \geq 0.1 M_\odot \times \left(\frac{Z}{Z_\odot}\right)^{1.5}\, .\label{eq:cb}
    \end{equation}
    The positive scaling with $Z$ quantitatively reproduces the observed data for close binaries, while the normalization ($0.1 M_\odot$) allows for stellar-mass secondaries to be produced. Disks that form such binaries will then be completely disrupted and planet formation is avoided.      
    \item Around stars that can do not harbour these harmful binaries, 
    Kepler planets are formed if 
    \begin{equation}
        M_{\rm solid} \geq 30 M_\oplus \, ,
        \label{eq:kepler}
    \end{equation}

    \item while disks with larger solid masses can continue to form Jovian planets. We set this to be
    \begin{equation}
        M_{\rm solid} \geq 150 M_\oplus\, .
        \label{eq:jovian}
    \end{equation}
    These choices reproduce the observed occurrences: $\sim 30\%$ for Kepler planets, and a third of that for giant planets. And they ensure that giant planets only occur in systems with inner {\it Kepler} planets, as is observed.
\end{itemize}
  
The results of such a model are presented in Fig. \ref{fig:scat}. While the left panel gives a visual impression of where the three types of objects fall in the plane of disk mass versus metallicity, the right panel shows quantitative comparison against observations. The agreements are good, with the only exception being the occurrence rate of Jovian planets at the lowest metallicity bin. However, the observed value here is suspicious --- the three metallicity bins do not exhibit a monotonic behaviour. 

The success of our simple model is somewhat surprising. We have a total of $6$ parameters: $4$ for the birth conditions, and $2$ for the disk mass distribution. In comparison, there are $9$ data points (Fig. \ref{fig:scat}). Moreover, we have not considered a slew of physical processes that may affect the outcome. The success likely stems from the fundamental role of solid mass for planet formation.

Assuming this model is correct, we can draw a few conclusions:
\begin{itemize}
    
    \item For the metallicity distribution that we adopt, mean metallicites are: close binary, $[Fe/H]=-0.003$; {\it Kepler} planets, $[Fe/H]=0.04$; giant planets, $[Fe/H]=0.10$. 
    
    \item While the fraction of close binaries continue to rise with decreasing metallicity, by comparison, at $Z \sim 0.5 Z_\odot$ (${\rm [Fe/H]} \sim -0.3$), the Kepler planet fraction has dropped by a factor of few to $\sim 7\%$. This reduction is more extreme for Jovian planets. These predictions can be tested by TESS, GAIA, PLATO or other large surveys.

    \item The solid mass requirement for {\it Kepler} planets, at $\sim 30 M_{ \oplus}$ is fairly reasonable: each {\it Kepler} planet likely contains some $10 M_{ \oplus}$, and each {\it Kepler} system contains, on average,  $3$ planets \citep{Zhu2018a}. If the observed disk masses are reliable, this suggests that Nature does not waste much when making these planets. 
    
    \item The solid mass requirement for giant planets has interesting implications. The correlation between cold Jovians and Kepler planets suggests that planets are built in a chain-like fashion, starting from the inside. So a higher solid mass may be required to extend the link to beyond an AU, where Jovian planets can potentially form.  Alternatively, the higher solid mass may boost the number density of planets on the chain, leading to earlier mergers and run-away gas accretion.
    
    \item Lastly, the factor of $5$ between equations (\ref{eq:kepler})-(\ref{eq:jovian}) is dependent on our assumed spread in disk mass.
    \end{itemize}

\section{Summary}\label{sec:discussion} 

Our main results are as follows: 

\begin{itemize}
    \item The sizes, and therefore masses, of \textit{Kepler} planets are not determined by stellar metallicity. The apparent correlation is a by-product of other known correlations.
    \item The occurrence rate of \textit{Kepler} planets  depends weakly (sublinearly) on stellar metallicity. This is related to stellar binaries and is not because these planets can't form in metal poor systems.
    \end{itemize}

The first result is anti-intuitive, since theories (\S \ref{sec:Intro}) tend to predict that more massive planets are produced in disks with more solids. There may be two ways why such an intuition is wrong. First, while metal-poor stars may indeed harbour disks with lower solid content, solid in the inner regions can be sourced from the entire disk (due, e.g., to grain drift or planetesimal migration) and its amount does not have to be pinned directly to the stellar metallicity. Second, there may be a characteristic mass at which generation-I planets are being produced. This mass may be related to disk properties such as gas scale heights, and is not related to the solid content. This latter point is supported by the finding in \citet{Wu2019}, where  the core sizes of {\it Kepler} planets are found to tightly correlate with the masses of host stars. 

The second result is equally surprising, when one contrasts it with the occurrence rate of giant planets  which rises strongly with stellar metallicity. {\it Kepler} planets have comparable masses as the cores of giant planets, and within the metallicity range that our data probe ($[Fe/H] \in [-0.3,0.3]$), their appearance is weakly affected by $Z$. When one accounts for systems of close binaries where planet formation is suppressed, the occurrence-$Z$ relation is further weakened. 

To resolve this conflict,  we create a crude phenomenological model. An essential ingredient in this model is a new variable:  the mass of the protoplanetary disk. This variable has a large spread and dilutes the metallicity dependence for Kepler planets, giving rise to the weak relation that we witness.  Giant planets, on the other hand, require a few times more solid mass than the Kepler planets do. And our model predicts that they should exhibit a much stronger metallicity dependence.
Meanwhile, when one extends our model to metallicity below $[Fe/H] \sim -0.3$, one expects the occurrence of Kepler planets to drop dramatically.

Overall, the formation of generation-I planets appears to be a fairly common outcome. For the limited metallicity range that we probe ($[Fe/H] \in [-0.3, 0.3]$), {\it Kepler} planets occur in some $30\%$ of all stars. If we include stars that are in close binaries, the total fraction rises to $\sim 50\%$. If so, what happens to the remaining $50\%$ stars that contain neither {\it Kepler} planets, nor close binaries?
Do they have too low disk masses?  Do they contain other, unknown population planets? Where are the elusive gen-II planets (one of which we are living on) being formed? Many important questions remain outstanding.

\section*{Acknowledgements}

We thank Wei Zhu, Maxwell Moe, Gwendolyn Eadie, and James Lane for helpful conversations, and NSERC for funding support. We thank our four referees for the many critical comments that helped  improve the paper. 

This work benefits from the NASA Exoplanet archive, and results from the NASA Kepler mission, the California Kepler Survey and the LAMOST survey. We also make use of the R package \textit{Mclust}. 




\appendix
\section{ The Apparent  $M_*-Z$ correlation for host stars}\label{ap:massmet}

The apparent correlation between  stellar mass and metallicity for planet hosts in the CKS sample (panel a in Fig. \ref{fig:GMM}) was first noted by \citet{Fulton2018}, who suggested that it arose because more massive stars are born later, out of a more polluted ISM. 
Indeed that panel (also see the figure below) shows an absence of massive, metal-poor stars. We had also adopted this view in our original manuscript, but criticisms from our reviewers prompt us to reconsider this stand. In this appendix, we show that this apparent correlation, strong as it is, is likely the result of sample selection. Fortunately, this change of stance does not affect any other conclusions we draw in the paper.

Multiple studies of stars in the solar neighbourhood have shown that stars born over the past $\sim 10$Gyrs have similar metallicity spread  \citep[e.g. ][]{Haywood2013,Bergemann2014,Holmberg}, with little  correlation between stellar mass and metallicity (see below).  Models of galactic chemical evolution have been proposed to explain this  \citep[e.g., `leaky box' model, see \S 8.5 of ][for review]{Pagel1997}. Moreover, 
the stars we are concerned with have a narrow mass range:  $0.7$ to $1.5M_\odot$. Given their long main-sequence lifetimes, we do not expect a significant difference between their mean ages. 

To understand our apparent $M_*-Z$ correlation, we adopt the sample of solar-neighbourhood stars from the Geneva-Copenhagen  \citep[][]{Nordstrom2004,Casagrande2011}, with refined parameters from the GALAH survey \citep[][]{Delgado2017, Buder2019}. 
By narrowing the GALAH sample to include only the mass and metallicity ranges spanned by the CKS stars ($ 0.7 <M_*< 1.6 $ and $ -0.47 <\text{[Fe/H]}< 0.44 $), and repeatedly drawing $692$ stars (the size of the CKS sample), we confirm that there is no significant correlation between $M_*$ and $Z$: the mean  mean correlation coefficient and p-value are $0.11$ and $0.03$, respectively. 

\begin{figure*}
    \centering
    \includegraphics[width = 0.51\textwidth]{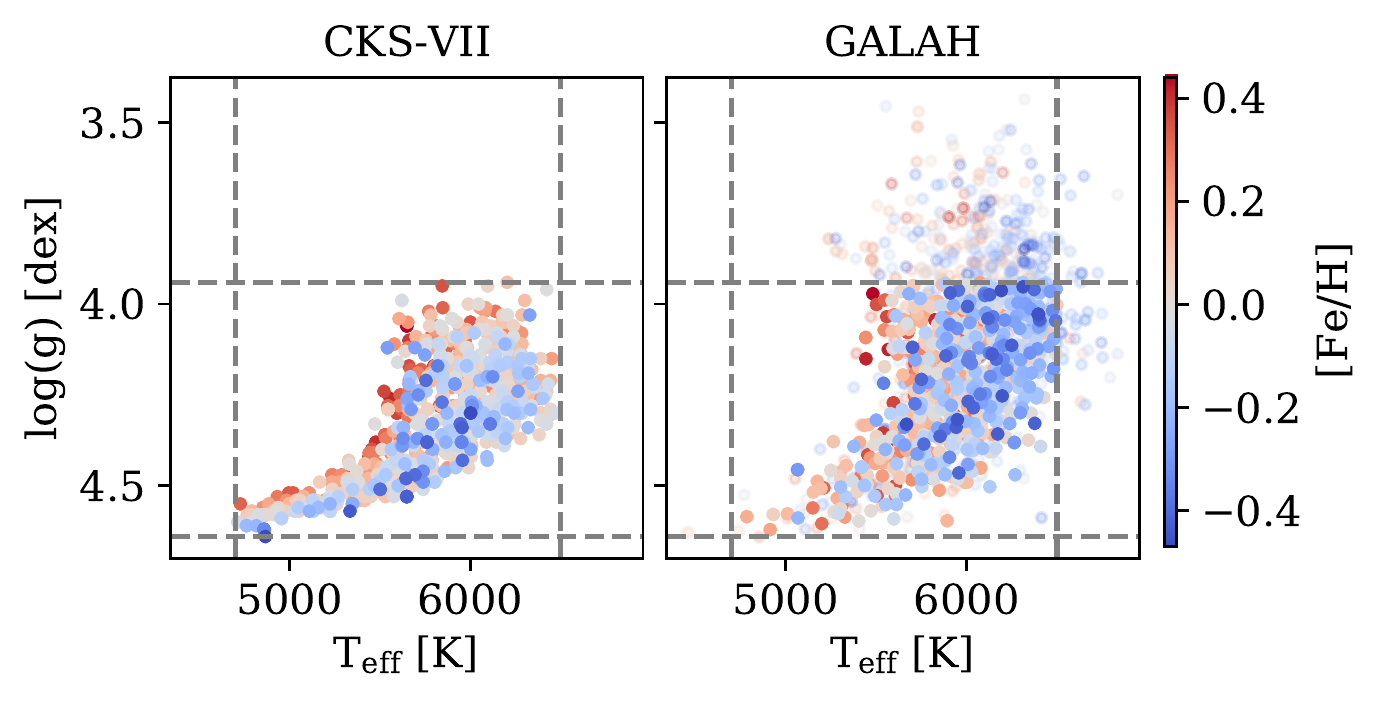}
    \includegraphics[width = 0.44\textwidth]{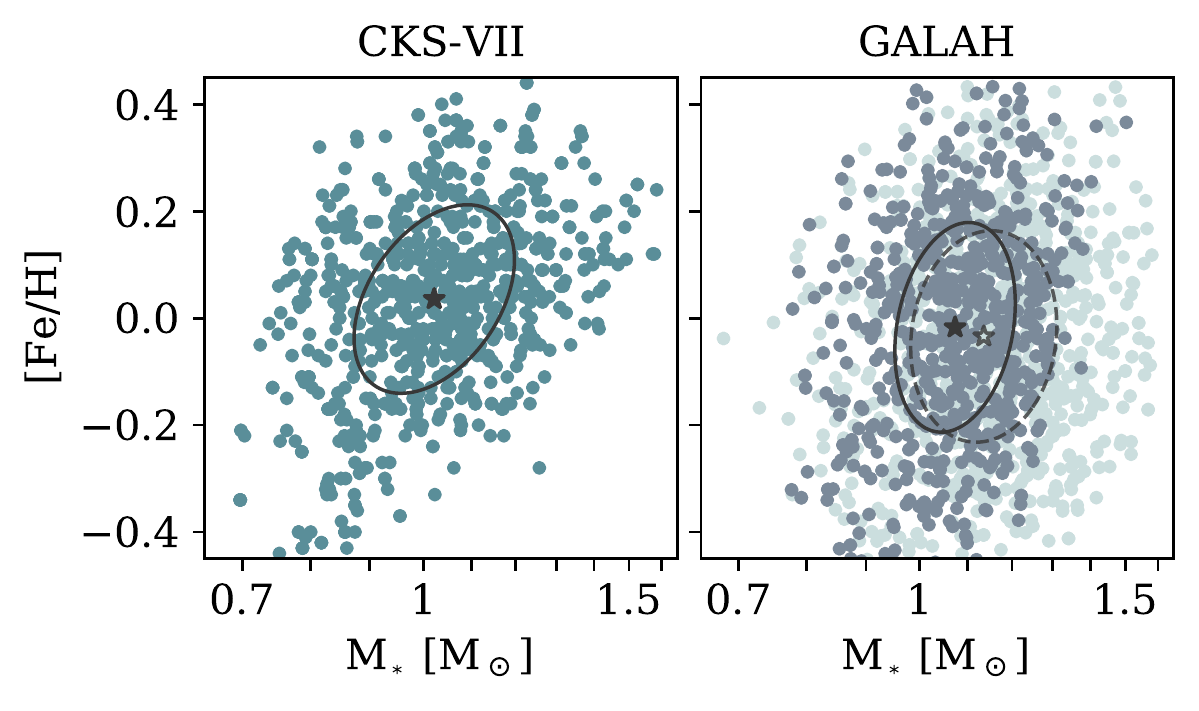}
    \caption{Experiment to illustrate that CKS sample selection gives rise to the apparent correlation between stellar mass and stellar metallicity. Limiting ourselves to stars with the same mass and metallicity range as the CKS host stars, and then applying the same cuts in $T_{\rm eff}$ and $\log(g)$ as are done for the CKS sample (left-most panel, four dashed lines), we obtain a biased GALAH sample (second panel from the left, showing 1000 stars, and those excluded ones as more transparent). The two right panels display the stars in the mass-metallicity plane, with the covariance illustrated by the ellipses. While the original GALAH sample show no correrlation (dashed ellipse), the selected one do.
    }
    \label{fig:logg-teff}
\end{figure*}

The CKS team reported the following cuts to obtain their host sample:  $4700 {\rm K} < T_{\rm eff} < 6500 {\rm K}$ and $3.9< \log(g) < 5.9$ \citep{Petigura2018}. However, the left panel of Fig. \ref{fig:logg-teff} shows that their sample occupy an even narrower phase space : $3.94\leq \log(g)\leq 4.64$.  Applying  the same $T_{\rm eff}$ cut and the latter $\log(g)$ cut to the GALAH sample excludes a large number of stars, especially those that are more massive and more metal poor. As a result, the remaining sample now exhibit a strong correlation with $\rho = 0.22$ and $p=10^{-7}$. Moreover, the apparent trend of $Z \propto M_*^{0.88 \pm 0.12}$ is  consistent with that for the CKS sample (eq. \ref{eq:starz}, $Z \propto M_*^{1.08\pm 0.09}$). This leads us to suggest that sample selection is responsible for the apparent correlation between stellar mass and stellar metallicity.

\bibliographystyle{AASJournal}
\bibliography{refs} 
\end{document}